\documentclass[11pt]{article}
\usepackage[a4paper]{geometry}
\usepackage[font=footnotesize,labelfont=bf]{caption}
\usepackage{fancyhdr}
\usepackage{subcaption}
\usepackage{placeins}
\usepackage{amsmath}
\usepackage{amsfonts}
\usepackage{graphicx,psfrag,epsf}
\usepackage{enumerate}
\usepackage[linesnumbered,ruled]{algorithm2e}
\usepackage{tabu}
\usepackage{setspace}
\usepackage{tikz}
\usepackage{float}
\usepackage{bm}
\usepackage{color}
\usepackage{listings}
\usepackage[square,sort,comma,numbers]{natbib}
\usepackage[hidelinks]{hyperref}
\usepackage[open]{bookmark}
\usepackage{booktabs}
\usepackage{rotating}
\usepackage{authblk}

\SetArgSty{textnormal}
\SetKwInput{KwTrain}{Train}

\fancypagestyle{firststyle}
{
	\fancyhf{}
	\fancyfoot[L]{\fontsize{9}{11}\selectfont Accepted for publication in \textit{Frontiers in Neuroscience}, vol. 16, no. 830630, at \url{https://doi.org/10.3389/fnins.2022.830630}. \newline \copyright \ 2022. This manuscript version is made available under the CC-BY 4.0 license \url{https://creativecommons.org/licenses/by/4.0/}}
}

\begin{document}

\title{Analyzing hierarchical multi-view MRI data with StaPLR: An application to Alzheimer's disease classification}
\date{February 28, 2022}
\author[1]{Wouter van Loon}
\author[1,2,3]{Frank de Vos}
\author[1]{Marjolein Fokkema}
\author[4,5]{Botond Szabo}
\author[6]{Marisa Koini}
\author[6]{Reinhold Schmidt}
\author[1,3]{Mark de Rooij}
\affil[1]{Department of Methodology and Statistics, Leiden University}
\affil[2]{Department of Radiology, Leiden University Medical Center}
\affil[3]{Leiden Institute for Brain and Cognition}
\affil[4]{Department of Decision Sciences, Bocconi University}
\affil[5]{Bocconi Institute for Data Science and Analytics, Bocconi University}
\affil[6]{Division of Neurogeriatrics, Department of Neurology, Medical University of Graz}

\maketitle

\thispagestyle{firststyle}

\begin{abstract}
	\noindent Multi-view data refers to a setting where features are divided into feature sets, for example because they correspond to different sources. Stacked penalized logistic regression (StaPLR) is a recently introduced method that can be used for classification and automatically selecting the views that are most important for prediction. We introduce an extension of this method to a setting where the data has a hierarchical multi-view structure. We also introduce a new view importance measure for StaPLR, which allows us to compare the importance of views at any level of the hierarchy. We apply our extended StaPLR algorithm to Alzheimer's disease classification where different MRI measures have been calculated from three scan types: structural MRI, diffusion-weighted MRI, and resting-state fMRI. StaPLR can identify which scan types and which derived MRI measures are most important for classification, and it outperforms elastic net regression in classification performance.
	
\end{abstract}

\textbf{keywords} \textit{multimodal MRI}, \textit{machine learning}, \textit{stacked generalization}, \textit{penalized regression}, \textit{feature selection}

\newpage

\section{Introduction}

In biomedical research, the integration of data from different sources into a single classification model is becoming increasingly common \citep{multiview_bio, fratello2017}. This is fueled by the increasing availability of multi-source data, for example through the UK Biobank \citep{UKbiobank, UKbiobank_imaging}, the Alzheimer's Disease Neuroimaging Initiative (ADNI) \citep{ADNI}, and various dementia registries around the world \citep{dementia_registries} such as the Prospective Registry on Dementia (PRODEM) \citep{PRODEM}. Training a model on multiple data sources has been found to increase accuracy in the prediction of brain-age \citep{Liem2017} and the classification of Alzheimer's disease (AD) \citep{Schouten2016, Rahim2016, Li2011}. \par 
A general term for data in which the features are divided into feature sets (for example, by source or modality) is \textit{multi-view data}, and the field of developing algorithms for such data is known as \textit{multi-view (machine) learning} \citep{multiview_review, multiview_book}. Of particular interest to this study is the multi-view learning framework known as \textit{multi-view stacking} \citep{Li2011, multiview_stacking, StaPLR}. The general idea of multi-view stacking is to first train a model on each feature set (also called a \textit{view}) separately. Then, each of these models is cross-validated to obtain a set of predictions of the outcome. Finally, another algorithm (called the \textit{meta-learner}) is trained on these cross-validated predictions. The meta-learner thus learns how to best combine the predictions from the view-specific models. Several methods used in previous neuroscience studies can be considered a form of multi-view stacking, and have shown better performance than single-view or non-stacked approaches \citep{Li2011, deVos2016, Rahim2016, Liem2017, Salvador2019, Guggenmos2020, Engemann2020, Ali2021}. However, these methods are generally ad-hoc approaches tailored specifically to the data at hand, and there is little consistency between the used methods. \par
Although previous studies have mostly focused on improving classification accuracy, it is also important to identify which views are relevant for prediction. For example, if a certain scan modality turns out to be irrelevant for prediction of a disease, it may not have to be measured at all. Recently, a variant of multi-view stacking called \textit{stacked penalized logistic regression} (StaPLR) has been developed specifically for this purpose \citep{StaPLR}. StaPLR essentially integrates the penalized logistic regression models which are already commonly used in neuroimaging classification, such as ridge regression \citep{ridge, ridge_logistic} and the lasso \citep{lasso, glmnet}, into a single unified multi-view stacking methodology. StaPLR can be used to select the feature sets that are most relevant for prediction, and has been shown to have several advantages over earlier methods, including a decreased false positive rate in view selection and a large reduction in computation time, while maintaining good classification accuracy \citep{StaPLR}. StaPLR is, to our knowledge, the only multi-view learning method which can be extended to hierarchical multi-view structures with an arbitrary number of levels while keeping computational feasibility. By hierarchical multi-view data we mean that feature sets are nested in other feature sets. Consider, for example, data collected from different domains, such as genetics, neuroimaging, and cognitive tests. Each of these domains could be considered a different view of the patients under consideration. These views could then be further divided into subviews. For example, the higher-level neuroimaging feature set could be further divided into lower-level feature sets corresponding to different scan types. \par 
In this study we propose an extension of the StaPLR method to hierarchical multi-view data. We show an application of this extension to an Alzheimer's disease classification problem based on three MRI scan types: structural MRI, diffusion-weighted MRI, and resting state functional MRI. For each of these scan types, different MRI measures were computed, where each measure is represented by multiple features. This yields a hierarchical multi-view structure with three levels: the \textit{features} (base level) are nested in the \textit{MRI measures} (intermediate level), which in turn are nested in the different \textit{scan types} (top level). Parts of this multi-view data set, which consists of data collected as part of PRODEM \citep{PRODEM} and the Austrian Stroke Prevention Study (ASPS) \citep{ASPS1, ASPS2}, have been used in previous studies \citep{deVos2017, Schouten2016, Schouten2017}, but this is the first time these features are all included into a single analysis. Previous applications of StaPLR have focused solely on a setting with two levels \citep{StaPLR, StaPLR2}. In this paper we therefore extend StaPLR to the hierarchical structure of the data. We will show how StaPLR can be used to both perform classification and identify the views that are most important for prediction. To provide a `benchmark' for the classification performance and interpretability of the model we additionally perform logistic elastic net regression \citep{elasticnet}, which has been used in many previous multi-view neuroimaging classification studies \citep{Trzepacz2014, Teipel2015, Bowman2016, Nir2016, Schouten2016, deVos2016}. We also compare the proposed extension with the original StaPLR algorithm. \par 
In addition to its advantages in view selection and computation time \citep{StaPLR}, the proposed extension of StaPLR has important advantages in terms of the interpretability of the resulting classifier. First, measures of view and feature relevance are readily available in the form of coefficients in a logistic regression model. This is in contrast to previous multi-view stacking methods focused on prediction accuracy, such as those using random forests as a meta-learner \citep{Liem2017, Engemann2020}. Second, extending StaPLR to match the hierarchical multi-view structure of the data allows us to calculate such measures of importance \textit{at each level of the hierarchy}. Thus, we can obtain estimates of the contribution of each scan type, but also of each MRI measure within those scan types. Finally, we show in section \ref{sect:mrm} how the proposed extension of StaPLR allows us to compare the contribution of different MRI measures even if they correspond to different scan types. \par
The application to the current data set aims to provide an example of a more general class of applications within neuroimaging and biomedical science as a whole. Since our focus is on demonstrating the methodology rather than on the specific data set, we will refrain from any interpretation regarding the specific clinical meaning of our findings with respect to the target population.

\section{Materials and methods}

\subsection{Participants}

Our data set consisted of 76 patients clinically diagnosed with probable AD, and 173 cognitively normal elderly controls, for a total of 249 observations. The AD patients were scanned at the Medical University of Graz as part of PRODEM \citep{PRODEM}. The elderly controls were scanned at the same scanning site, with the same scanning protocol, and over the same time period as part of the ASPS \citep{ASPS1, ASPS2}. We only included patients for which anatomical MRI, diffusion MRI and rs-fMRI were available. 

\FloatBarrier

\subsection{MRI analysis}
The scanning protocols, and an elaborate description of the MRI analyses are provided in the supplementary materials. For each scan type, several MRI measures were computed; below we provide a brief description. An overview of the features included in our analyses is presented in Table \ref{tab:features}. \par 
The strutural MRI scans were used to calculate five different MRI measures. Grey matter density refers to the percentage of grey matter in a certain area of the brain. The 48 features correspond to the 48 regions of the probabilistic Harvard-Oxford cortical atlas \citep{HOA}. Subcortical volumes describe the size of several subcortical brain structures. The 14 features correspond to the thalamus, caudate, putamen, pallidum, hippocampus, amygdala and accumbens of both hemispheres. The neocortex was parcellated into the 68 regions of the Desikan-Killiany atlas \citep{Desikan2006}. For each of these regions, the mean cortical thickness, mean cortical curvature, and the total area of the region's cortical surface (``cortical area'') was calculated. \par
The diffusion-weighted MRI scans were used to calculate fractional anisotropy, mean diffusivity, axial diffusivity, and radial diffusivity for the 20 white matter regions of the JHU white-matter tractography atlas \citep{Hua2008}. \par 
The resting state fMRI scans were used to calculate multiple types of functional connectivity (FC) estimates. First, temporal concatenation independent component analysis (ICA) was used to extract both 20 and 70 components \citep{deVos2017}. For both of these configurations, FC matrices were calculated using either full or sparse partial correlations, resulting in four different FC matrices for each participant. These four matrices were further used to calculate FC dynamics using a sliding window approach. The FC matrices were calculated for each time window, and the standard deviation of these matrices over time reflect the FC dynamics. In addition, the sliding window matrices of all participants were clustered into five connectivity states using k-means clustering. The number of sliding window matrices assigned to each of these five states was calculated for each participant. The four FC matrices were also used to calculate several common graph metrics. Additionally, voxel-wise connectivity with 10 different resting state networks was calculated using dual regression, as well as seed based connectivity with both the left and right hippocampus as seed regions. Furthermore, a voxel-wise eigenvector centrality map was calculated. Eigenvector centrality attributes a value to each voxel in the brain such that a voxel receives a large value if it is strongly correlated with many other voxels that are themselves central within the network. Lastly, the amplitude of low frequency fluctuations (ALFF), and its weighted variant the fractional ALFF (fALFF), were calculated for each voxel. Details can be found in the supplementary materials and in \citet{deVos2016}.

\begin{table}
	\centering
	\resizebox{\textwidth}{!}{
		\begin{tabular}{llr}
			\hline
			($s$) scan type & ($v$) MRI measure & number of features \\
			\hline
			(1) structural MRI & (1) grey matter density & 48 \\
			& (2) subcortical volumes & 14 \\
			& (3) cortical thickness & 68 \\
			& (4) cortical area & 68 \\
			& (5) cortical curvature & 68 \\
			\hline
			(2) diffusion MRI & (6) fractional anisotropy & 20 \\
			& (7) mean diffusivity & 20 \\
			& (8) axial diffusivity & 20 \\
			& (9) radial diffusivity & 20 \\
			\hline
			(3) resting state fMRI & (10) full FC correlation matrix (20 $\times$ 20) & 190 \\
			& (11) full FC correlation matrix (70 $\times$ 70) & 2,415 \\
			& (12) sparse partial FC correlation matrix (20 $\times$ 20) & 190${}^{[*]}$ \\
			& (13) sparse partial FC correlation matrix (70 $\times$ 70) & 2,415${}^{[*]}$ \\
			& (14) SD of full FC matrix (20 $\times$ 20) & 190 \\
			& (15) SD of full FC matrix (70 $\times$ 70) & 2,415 \\
			& (16) SD of sparse partial FC matrix (20 $\times$ 20) & 190 \\
			& (17) SD of sparse partial FC matrix (70 $\times$ 70) & 2,415${}^{[*]}$ \\
			& (18) FC states of full FC matrix (20 $\times$ 20) & 5 \\
			& (19) FC states of full FC matrix (70 $\times$ 70) & 5 \\
			& (20) FC states of sparse partial FC matrix (20 $\times$ 20) & 5 \\
			& (21) FC states of sparse partial FC matrix (70 $\times$ 70) & 5 \\
			& (22) Graph metrics of full FC matrix (20 $\times$ 20) & 124 \\
			& (23) Graph metrics of full FC matrix (70 $\times$ 70) & 424 \\
			& (24) Graph metrics of sp. par. FC matrix (20 $\times$ 20) & 124${}^{[*]}$ \\
			& (25) Graph metrics of sp. par. FC matrix (70 $\times$ 70) & 424${}^{[*]}$ \\
			& (26) FC with visual network 1 & 190,981 \\
			& (27) FC with visual network 2 & 190,981 \\
			& (28) FC with visual network 3 & 190,981 \\
			& (29) FC with default mode network & 190,981 \\
			& (30) FC with the cerebellum & 190,981 \\
			& (31) FC with sensorimotor network & 190,981 \\
			& (32) FC with auditory network & 190,981 \\
			& (33) FC with executive control network & 190,981 \\
			& (34) FC with frontoparietal network 1 & 190,981 \\
			& (35) FC with frontoparietal network 2 & 190,981 \\
			& (36) FC with left hippocampus & 190,981 \\
			& (37) FC with right hippocampus & 190,981 \\
			& (38) Fast eigenvector centrality mapping & 190,981 \\
			& (39) ALFF & 190,981 \\
			& (40) fALFF & 190,981 \\
			\hline
			total & & 2,876,515${}^{[**]}$\\
			\hline
	\end{tabular}}
	\caption{Overview of the scan types, MRI measures, and corresponding number of features used in this study. The indices $s$ and $v$ are used to refer to the different scan types and MRI measures in Algorithm \ref{al:1}. ${}^{[*]}$Some of these features were removed due to having a variance of zero; the total number of features after removal for each of these MRI measures is 189, 2337, 2414, 123 and 423, respectively. ${}^{[**]}$The removal of features due to zero variance is already reflected in this total.}  \label{tab:features}
\end{table}

\FloatBarrier

\subsection{Logistic elastic net regression (benchmark)}

To provide a reference value for the accuracy and area under the receiver operating characteristic curve (AUC) we performed logistic elastic net regression \citep{elasticnet}. Elastic net regression employs a mixture of $L_1$ and $L_2$ penalties on the vector of regression coefficients \citep{elasticnet}. The $L_1$ penalty can perform feature selection by setting some coefficients to zero, while the $L_2$ penalty encourages groups of correlated features to be selected together. Elastic net regression operates at the level of the features and thus ignores the multi-view structure of the data completely. Elastic net regression has two tuning parameters, one which determines the mixture of $L_1$ and $L_2$ penalties ($\alpha$), and one which determines the amount of penalization ($\lambda$). We selected a value for both penalties through 10-fold cross-validation. For $\alpha$, the set of candidate values is a sequence from 0 to 1 in increments of 0.1. The set of candidate values for $\lambda$ is a sequence of 100 values adaptively chosen by the software \citep{glmnet}. In particular, the 100 values are decreasing values on a log scale from $\lambda_{\text{max}}$ to $\lambda_{\text{min}}$, where $\lambda_{\text{max}}$ is the smallest value such that the entire regression coefficient vector is zero, and $\lambda_{\text{min}} = \epsilon \lambda_{\text{max}}$ \citep{glmnet}. We set $\epsilon = 0.01$ (the default). To prevent overfitting, we assessed the models classification performance using a double (nested) cross-validation approach \citep{Varma2006}: an inner loop is used to determine the values of the tuning parameters, and an outer loop is used for determining classification accuracy and AUC. For both the inner and outer loop we used 10 folds. Additionally, we repeat this nested cross-validation approach 10 times to average out the effects of different allocations of the subjects to the folds. Elastic net regression was performed in R 4.0.2 \citep{Rcore}, using the package \texttt{glmnet} 1.9-8 \citep{glmnet}. \par 
Since the elastic net ignores the multi-view structure of the data, it is hard to infer the importance of an MRI measure or scan type. After all, a single MRI measure can be represented by anywhere from 5 to over 190,000 regression coefficients. With a total of over 2.8 million features, showing the results for each feature individually is infeasible. In order to summarize the results at the MRI measure level we calculated for each measure the following: (1) the number of non-zero coefficients, and (2) the $L_2$-norm (i.e. the square root of the sum of squared values) of the associated regression coefficient vector.

\FloatBarrier

\subsection{Stacked penalized logistic regression}

From each of the three scan types several MRI measures are derived. In turn, each MRI measure consists of multiple features, as shown in Table \ref{tab:features}. Thus, the data have a hierarchical multi-view structure with three levels, and we therefore propose an extension of the StaPLR algorithm to three levels. The previous version of StaPLR only allowed for a two-level structure, where features were nested within views \citep{StaPLR}. This means one has to choose to either use the MRI measures as views, or the scan types. Thus, one would have to either ignore part of the hierarchy, or perform two separate analyses. Algorithm \ref{al:1} presents the extension of StaPLR to 3 levels: We start by training a logistic ridge regression model on each of the 40 MRI measures under consideration (line 1). For each of these models we use 10-fold cross-validation to choose an appropriate value for the penalty parameter. The reason we use ridge regression at this step is that we are not interested in selecting individual features, only entire MRI measures. We refer to the classifiers that were obtained for each of the 40 MRI measures as $\hat{f}_1, \dots, \hat{f}_{40}$. Since these classifiers are probabilistic, they give predicted values in $[0,1]$. \par 
For each scan type $s$, we want to obtain an intermediate classifier $\hat{f}_{\text{inter}}^{(s)}$ that combines the predictions of the classifiers trained on the corresponding MRI measures. For example, for the structural MRI scan type, we want to obtain an intermediate classifier $\hat{f}_{\text{inter}}^{(1)}$ that combines the predictions of  $\hat{f}_1$ through $\hat{f}_5$, which are the classifiers corresponding to grey matter density, subcortical volumes, cortical thickness, cortical area, and cortical curvature. In order to train such a combination model, we need a vector of predictions for each of the classifiers $\hat{f}_1$ through $\hat{f}_5$. 
We could simply use the fitted values for each of these classifiers, but this would yield overly optimistic estimates of predictive accuracy, because the same data would be used for fitting the model and generating predictions. Instead, we would like to obtain a vector of estimated out-of-sample predictions \citep{Wolpert1992}. We obtain such estimates through 10-fold cross-validation (line 2). We divide the observations into 10 folds, train each classifier on 9 folds, then generate predictions for the observations in the left-out fold. We repeat this procedure to obtain predictions for each of the folds. Note that ``training the classifier" includes the selection of penalty parameters; the cross-validation loop used to select the penalty parameter is nested within the loop used to generate the predictions. This means the predictions are truly made on data which the classifier has never seen. \par 
Once we obtained a vector of cross-validated predictions for each of the 40 classifiers, we collect them into 3 separate matrices, one for each scan type (line 3). These matrices then become the training data for the next step in the hierarchical StaPLR algorithm, where we train a nonnegative logistic lasso model on each of the 3 matrices of predictions (line 4). We apply the nonnegative lasso at this step because we would like to select a subset of the available MRI measures. The nonnegativity constraints have previously been shown to improve performance; see \citet{StaPLR} for empirical evidence and theoretical support.
We end up with 3 intermediate classifiers, one for each scan type. \par 
In order to train the meta-learner, we need to obtain a vector of estimated out-of-sample predictions for each of the 3 intermediate classifiers. We again do this using 10-fold cross-validation (line 5). We then collect these in another matrix (line 6), and train another logistic nonnegative lasso model on this matrix (line 7). The model training is now complete, and the final stacked classifier can be used by applying the classifiers $\hat{f}_1, \dots, \hat{f}_{40}$ to the 40 MRI measures, aggregating their predictions for each scan type using the intermediate classifiers $\hat{f}^{(1)}_{\text{inter}}$, $\hat{f}^{(2)}_{\text{inter}}$, $\hat{f}^{(3)}_{\text{inter}}$, and then combining the output of each scan type using the meta-classifier $\hat{f}_{\text{meta}}$ (line 8). \par 
The model is again evaluated using double (nested) 10-fold cross-validation: In the outer validation loop, the entirety of Algorithm \ref{al:1} is applied to 90\% of the data, and the remaining 10\% is used only for the calculation of classification accuracy and AUC. This procedure was repeated 10 times. StaPLR was performed in R using the package \texttt{multiview} 0.3.1 \citep{multiview}, using the default optimization settings. The scripts used for model fitting and evaluation are available in a public code repository \citep{code_repo}. For a more general discussion of the original StaPLR algorithm with only 2 levels we refer to \citet{StaPLR}.

\begin{algorithm} 
	\caption{StaPLR with 3 levels, as applied to the current data set} \label{al:1}
	\DontPrintSemicolon
	\KwData{$\bm{X}^{(v)}$, $v=1 \dots 40$, the 40 different MRI measures as shown in Table \ref{tab:features}, and $\bm{y}$ the binary outcome variable, where a value of 1 indicates probable Alzheimer disease, and a value of 0 indicates a healthy control.}
	\BlankLine
	\BlankLine
	Train a logistic ridge regression classifier (including cross-validation for $\lambda$) on the pairs ($\bm{X}^{(v)}, \bm{y}$), $v = 1, \dots, 40$, to obtain view-specific classifiers $\hat{f}_1, \dots, \hat{f}_{40}$. \;
	Apply 10-fold cross-validation to obtain a vector of predictions $\bm{z}^{(v)}$ for each of the $\hat{f}_v$, $v = 1, \dots, 40$. \;
	For each of the three scan types $s = 1, 2, 3$, collect the predictions $\bm{z}^{(v)}$ which correspond to that scan type column-wise into the matrix $\bm{Z}^{(s)}$. \;
	Train a logistic nonnegative lasso classifier (including cross-validation for $\lambda$) on the pairs ($\bm{Z}^{(s)}, \bm{y}$), $s = 1,2,3$, to obtain the intermediate classifiers $\hat{f}_{\text{inter}}^{(1)}$, $\hat{f}_{\text{inter}}^{(2)}$, $\hat{f}_{\text{inter}}^{(3)}$. \;
	Apply 10-fold cross-validation to obtain a vector of predictions $\bm{z}^{(s)}_{\text{inter}}$ for each of the $\hat{f}_{\text{inter}}^{(1)}$, $\hat{f}_{\text{inter}}^{(2)}$, $\hat{f}_{\text{inter}}^{(3)}$. \;
	Collect the predictions $\bm{z}^{(1)}_{\text{inter}}$, $\bm{z}^{(2)}_{\text{inter}}$, $\bm{z}^{(3)}_{\text{inter}}$ column-wise into the matrix $\bm{Z}_{\text{inter}}$. \;
	Train a logistic nonnegative lasso classifier (including cross-validation for $\lambda$) on the pair ($\bm{Z}_{\text{inter}}, \bm{y}$) to obtain a meta-classifier $\hat{f}_{\text{meta}}$. \;
	Define the final stacked classifier as: 
	\begin{equation*}
		\begin{aligned}
			\hat{f}_{\text{stacked}}(\bm{X}) \quad = \quad & \hat{f}_{\text{meta}}(\hat{f}_{\text{inter}}^{(1)}(\hat{f}_1(\bm{X}^{(1)}), \dots, \hat{f}_5(\bm{X}^{(5)})), \\
			& \hat{f}_{\text{inter}}^{(2)}(\hat{f}_6(\bm{X}^{(6)}), \dots, \hat{f}_9(\bm{X}^{(9)})), \\
			& \hat{f}_{\text{inter}}^{(3)}(\hat{f}_{10}(\bm{X}^{(10)}), \dots, \hat{f}_{40}(\bm{X}^{({40})}))).
		\end{aligned}
	\end{equation*}
	
\end{algorithm}

\FloatBarrier

\subsubsection{Quantifying feature set relevance across scan types} \label{sect:mrm}

One of the advantages of StaPLR is that at each level the method fits a logistic regression model. Thus, at each level, the classifiers can be interpreted as regular logistic regression models. This way one can easily determine the relative importance of the different scan types, or the different MRI measures within a scan type. However, if one wishes to compare feature sets corresponding to different scan types, for example an MRI measure from structural and one from functional MRI, an issue arises. Because the models used at each level are logistic regression models, and the logistic function is nonlinear, the final stacked classifier cannot be obtained by simply multiplying the regression weights of the different levels. If one wishes to compare the relative importance of feature sets across the different scan types, a different approach is needed. \par 
In hierarchical StaPLR, at the base level a separate classifier is trained on each MRI measure separately. Consider as an analogy a human committee: each base-level classifier can be considered a member of a committee providing a prediction of the outcome. The intermediate classifiers and meta-classifier then assign weights to the predictions of the committee members and combine them into a single predicted outcome. Now consider the possibility that one committee member makes a different prediction than all the others. In human committees such a dissenting opinion is sometimes called a \textit{minority report} \citep{dictionary, minority_report}. We can measure the impact of such a minority report by quantifying how the final predicted outcome changes as a single member changes its prediction, while the predictions of all the other members are kept constant. We will call this quantification the \textit{minority report measure} (MRM). Since in our case, each committee member is a classifier trained on a specific MRI measure, the MRM can be considered a measure of importance of this MRI measure in determining the final prediction. \par
The MRM measures the change in the outcome of the stacked classifier when the prediction corresponding to the $i$th MRI measure derived from scan type $s$ changes from value $a$ to value $b$, while all other predictions are kept constant at value $c$. Different choices for $a$, $b$ and $c$ are possible. In our analysis, we choose $a = 0$ and $b = 1$, which are the theoretical minimum and maximum, and $c = \bar{y}$, which is the proportion of observations corresponding to class 1. In this case the MRM measures the maximum possible change in final prediction attributable to the view $\bm{X}^{[s,i]}$, while the predictions corresponding to all other MRI measures are set to the sample mean of $y$. Other possible choices for $a$ and $b$ include the empirical minimum and maximum, respectively. \par 
In the context of StaPLR applied to the current data set, the MRM can be formalized as follows: Denote by $\bm{X}^{[s, i]}, i = 1 \dots m_s$, $s = 1 \dots S$, the $i$th MRI measure of scan type $s$, with $m_s$ the total number of measures corresponding to scan type $s$. Denote by $\hat{\beta}_0^{[s]}, s = 1 \dots S$ the intercept of the intermediate classifier corresponding to scan type $s$. Denote by $\hat{\beta}_i^{[s]}, i = 1 \dots m_s, s = 1 \dots S$ the coefficient of the $i$th MRI measure of scan type $s$. Denote by $\hat{\omega}_0$ the intercept of the meta-classifier, and by $\hat{\omega}^{[s]}, s=1 \dots S$ the weight of scan type $s$. 
Then for the $i$th MRI measure corresponding to scan type $s$, we define the MRM as:
\begin{equation}
	\begin{aligned}
		\text{MRM}(\bm{X}^{[s,i]}, a ,b, c) = g(\bm{X}^{[s,i]}, b, c) -  g(\bm{X}^{[s,i]}, a, c),	
	\end{aligned}
\end{equation}
with $a,b,c \in [0,1]$, $b > a$, and
\begin{equation}
	\begin{aligned}
		g(\bm{X}^{[s,i]}, b, c) = \psi \left( \hat{\omega}_0 + \hat{\omega}_s \psi \left( \hat{\beta}_0^{[s]} + \hat{\beta}_i^{[s]}b + \sum_{j \neq i} \hat{\beta}_j^{[s]} c \right ) + \sum_{k \neq s} \hat{\omega}_k \psi \left( \hat{\beta}_0^{[k]} + \sum_{j = 1}^{m_k} \hat{\beta}_j^{[k]} c \right) \right),
	\end{aligned}
\end{equation}
where $\psi$ denotes the logistic function, i.e.
\begin{equation}
	\psi(x) = \frac{\exp{(x)}}{1 + \exp{(x)}}.
\end{equation}
Note that, given $a$, $b$ and $c$, the value of the MRM depends only on the estimated parameters of the stacked model. The MRM can thus be readily calculated without any need for resampling or refitting of the model, unlike many model-agnostic measures of feature importance such as permutation feature importance \citep{Breiman2001, Fisher2019} or SHAP values \citep{SHAP}.

\FloatBarrier

\section{Results}

\subsection{Logistic elastic net regression}

\begin{figure}[]
	\centering
	\includegraphics{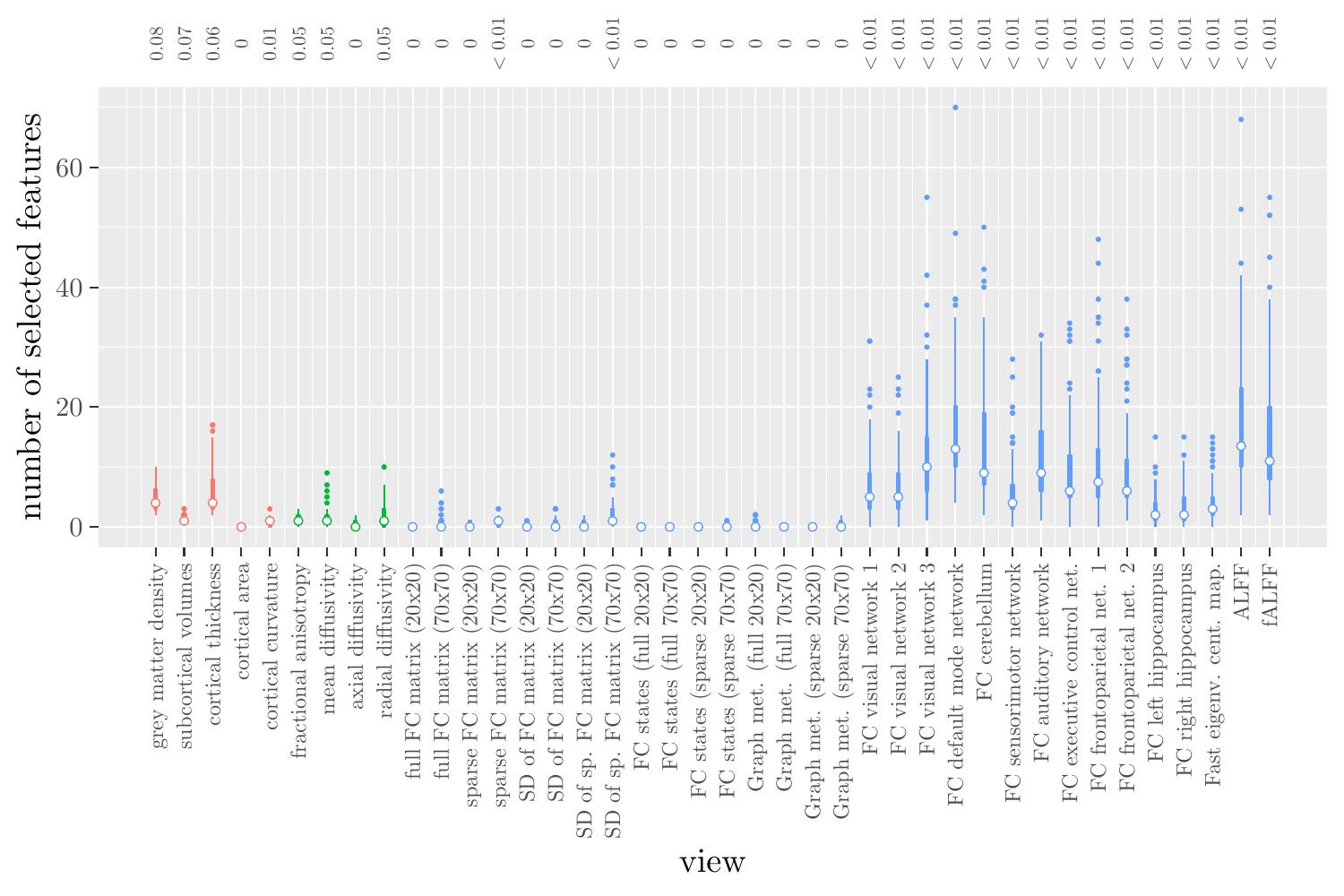}
	\caption{Boxplots of the number of selected features for each MRI measure resulting from the elastic net regression, colored by scan type (red = structural MRI, green = diffusion MRI, blue = resting state fMRI). The numbers at the top of the graph denote the median proportion of nonzero coefficients. \label{fig:elastic_plot}}
\end{figure}

The mean AUC of the model was 0.922 (SD = 0.008). The mean test accuracy of the model was 0.848 (SD = 0.012). The selected value of the tuning parameter $\alpha$ varied from 0.2 to 1, with an average of 0.788. On average, the model contained 168.17 (SD = 113.72) features. On average, the selected features were spread out over 24.07 (SD = 3.15) different views. Thus, elastic net regression provides classifiers which are fairly sparse at the feature level, but not at the MRI measure level. Consider Figure \ref{fig:elastic_plot}, which shows the distribution of the number of selected features for each MRI measure across the 10 $\times$ 10 fitted models. It can be observed that among the MRI measures with the largest median number of selected features are those which correspond to the voxel-wise functional connectivity with various RSNs (measures ($v$) 26 through 37). For all of these measures, the median number of selected features is greater than zero. It should be noted that these measures, along with ALFF and fALFF, are also those which contain by far the largest number of features. Each of these measures contains over 190,000 features, but the median number of selected features from each of them is typically around 5 to 15 (see also Figure \ref{fig:elastic_plot}). \par 
These results highlight several drawbacks of elastic net regression for multi-view data: elastic net regression tends to select a small number of features among a large number of MRI measures. This is not very useful from a data collection point of view, since one would typically collect or calculate an entire MRI measure. For example, one would have to perform the process of calculating the RSNs through ICA regardless of how many features were selected among measures 26 through 37. It is also not very useful from the viewpoint of model interpretation, since the functional connectivity of a resting state network with a handful of voxels scattered throughout the brain is unlikely to be very informative to a clinician. Additionally, comparing Figure \ref{fig:elastic_plot} with Table \ref{tab:features} shows that the MRI measures with the largest number of selected features are also the measures which contain the largest number of features to begin with. However, these are not necessarily the most important measures for predicting the outcome. Thus, given two views which are similarly predictive of the outcome, a view with a much larger number of features will likely have a much larger number of selected features. \par
\begin{figure}[]
	\centering
	\includegraphics{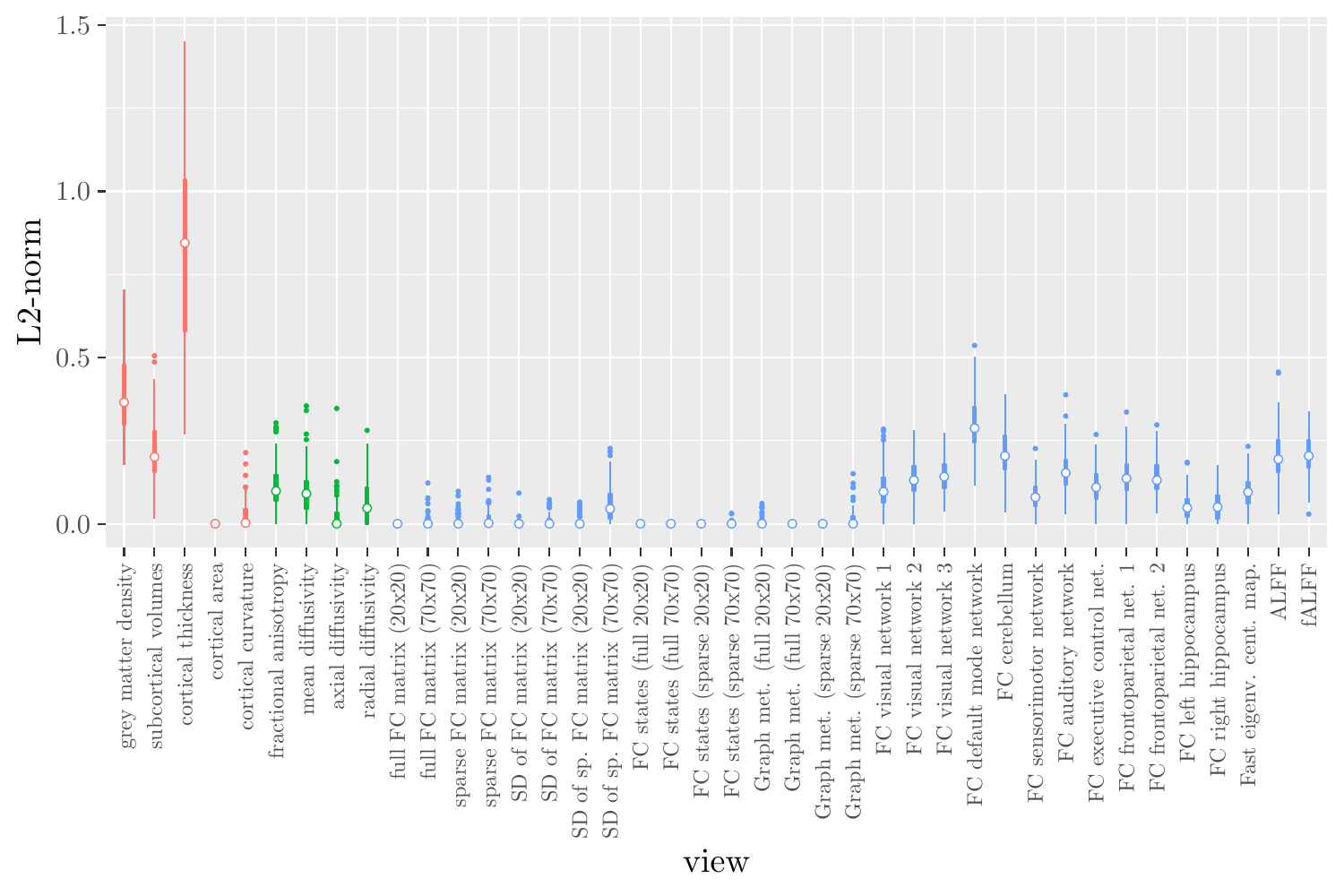}
	\caption{Boxplots of the $L_2$-norm of the regression coefficient vector for each MRI measure resulting from the elastic net regression, colored by scan type (red = structural MRI, green = diffusion MRI, blue = resting state fMRI). \label{fig:elastic_plot_norm}}
\end{figure}

 
Elastic net regression does not provide a direct measure of the importance of an MRI measure, since it operates at the feature rather than the view level. However, we can obtain a measure of the importance of an MRI measure by calculating the $L_2$-norm (i.e. the square root of the sum of squared values) of the corresponding regression coefficient vector. The results are shown in Figure \ref{fig:elastic_plot_norm}, where it can be observed that it is actually the structural MRI measures of grey matter density and cortical thickness which have the largest $L_2$-norm. Although Figures \ref{fig:elastic_plot} and \ref{fig:elastic_plot_norm} allow us to summarize the outcome at the MRI measure level, it is difficult to use the results of the elastic net regression to draw conclusions about which MRI measure is the most important for classification, or which MRI measures do not need to be measured in the future, since at least some features were selected from a large number of measures. Furthermore, in order to draw conclusions about the different scan types, one would have to re-aggregate the results at that level.

\FloatBarrier

\subsection{Original StaPLR algorithm}

The original StaPLR algorithm only allows for a two-level structure, with features nested within views. Thus one has to choose between using the MRI measures as views, or the scan types. Here we show the results of both choices. 

\subsubsection{MRI measures only}

\begin{figure}[]
	\centering
	\includegraphics{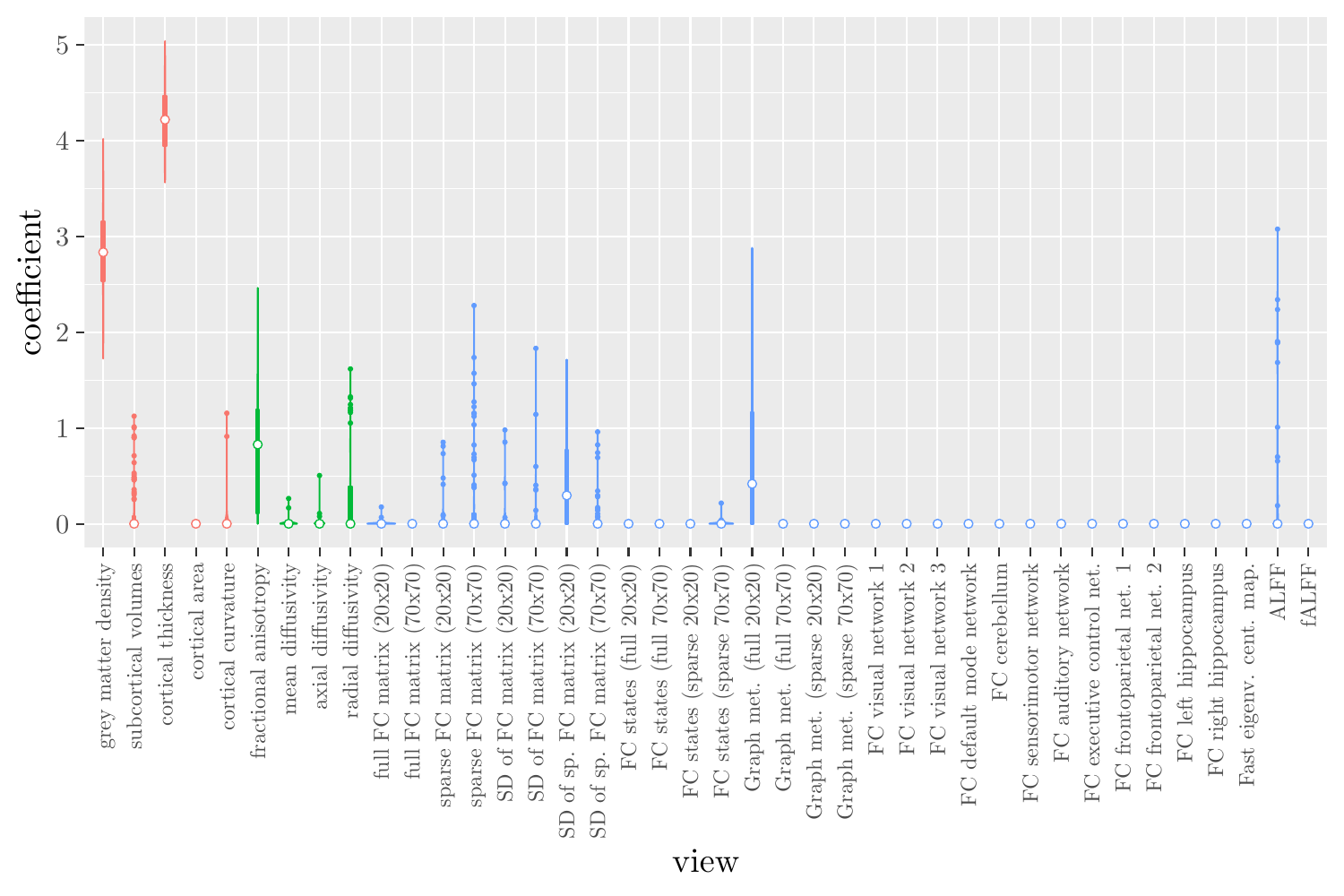}
	\caption{Boxplots of the regression coefficients of StaPLR applied with only 2 levels, using the MRI measures as views, colored by scan type (red = structural MRI, green = diffusion MRI, blue = resting state fMRI).
		\label{fig:MVS_bottom_coef_plot}}
\end{figure}

The mean AUC of the model using the MRI measures as views was 0.942 (SD = 0.006). The mean accuracy was 0.888 (SD = 0.007). The median regression coefficient for each MRI measure, as well as their distribution across the 10 $\times$ 10 fitted stacked classifiers can be observed in Figure \ref{fig:MVS_bottom_coef_plot}. The resulting classifier is considerably sparser than the one obtained through elastic net regression (Figure \ref{fig:elastic_plot_norm}). Note that this analysis only gives a measure of importance for each MRI measure, not for the scan types.

\subsubsection{Scan types only}

\begin{figure}[]
	\centering
	\includegraphics{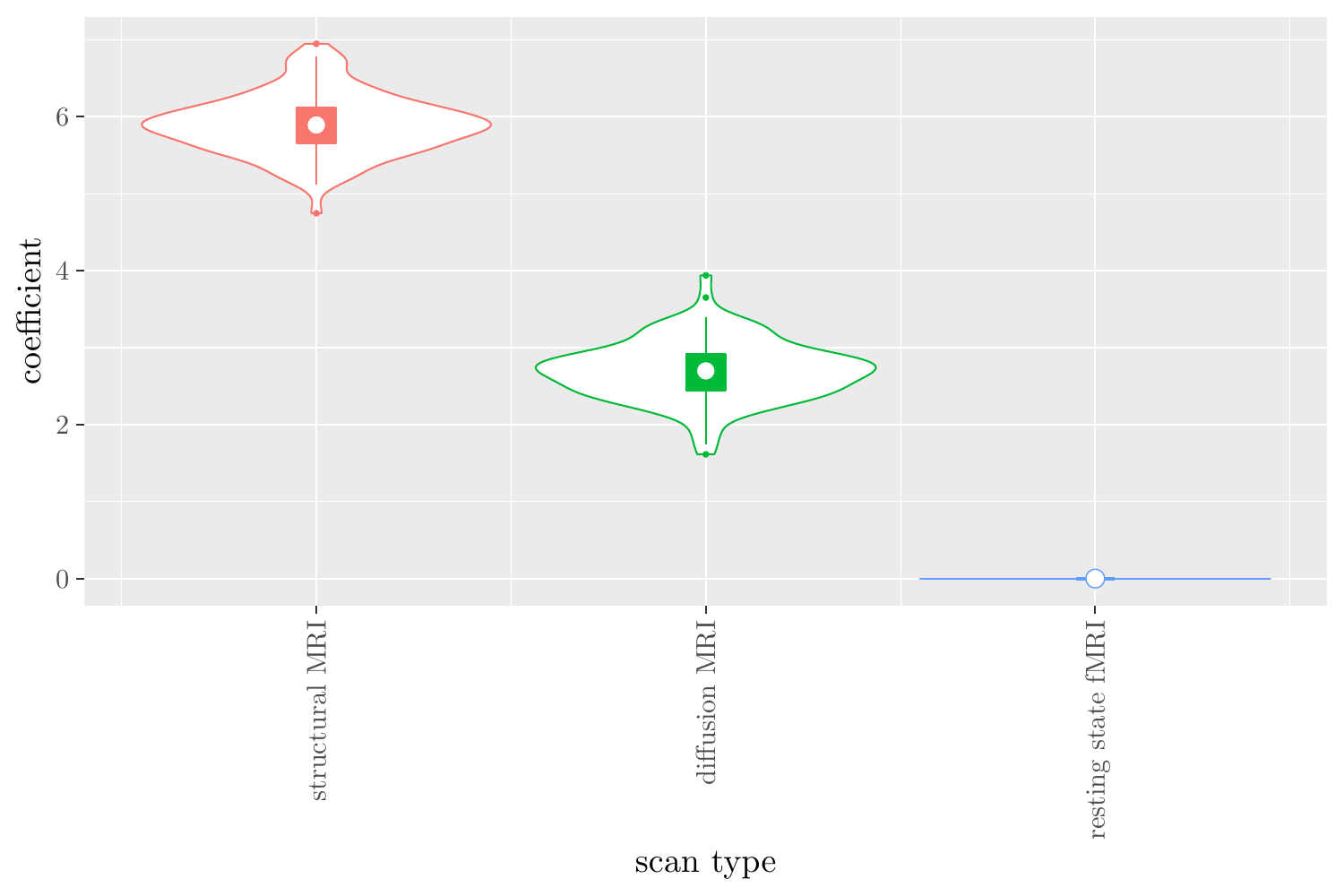}
	\caption{Box-and-violin plots of the regression coefficients of StaPLR applied with only 2 levels, using the scan types as views.
		\label{fig:MVS_top_coef_plot}}
\end{figure}

The mean AUC of the model using the scan types as views was 0.942 (SD = 0.008). The mean accuracy was 0.897 (SD = 0.006). The median regression coefficient for each scan type, as well as their distribution across the 10 $\times$ 10 fitted stacked classifiers can be observed in Figure \ref{fig:MVS_top_coef_plot}. Structural MRI obtains the highest coefficient, followed by diffusion MRI. Resting state fMRI is never selected. Note that this analysis does not perform selection of MRI measures, only of scan types. Thus, it is not possible to select a subset of relevant MRI measures for any scan type; the complete scan type has to be included or excluded from the model.

\FloatBarrier

\subsection{Hierarchical StaPLR}

\begin{figure}[]
	\centering
	\includegraphics{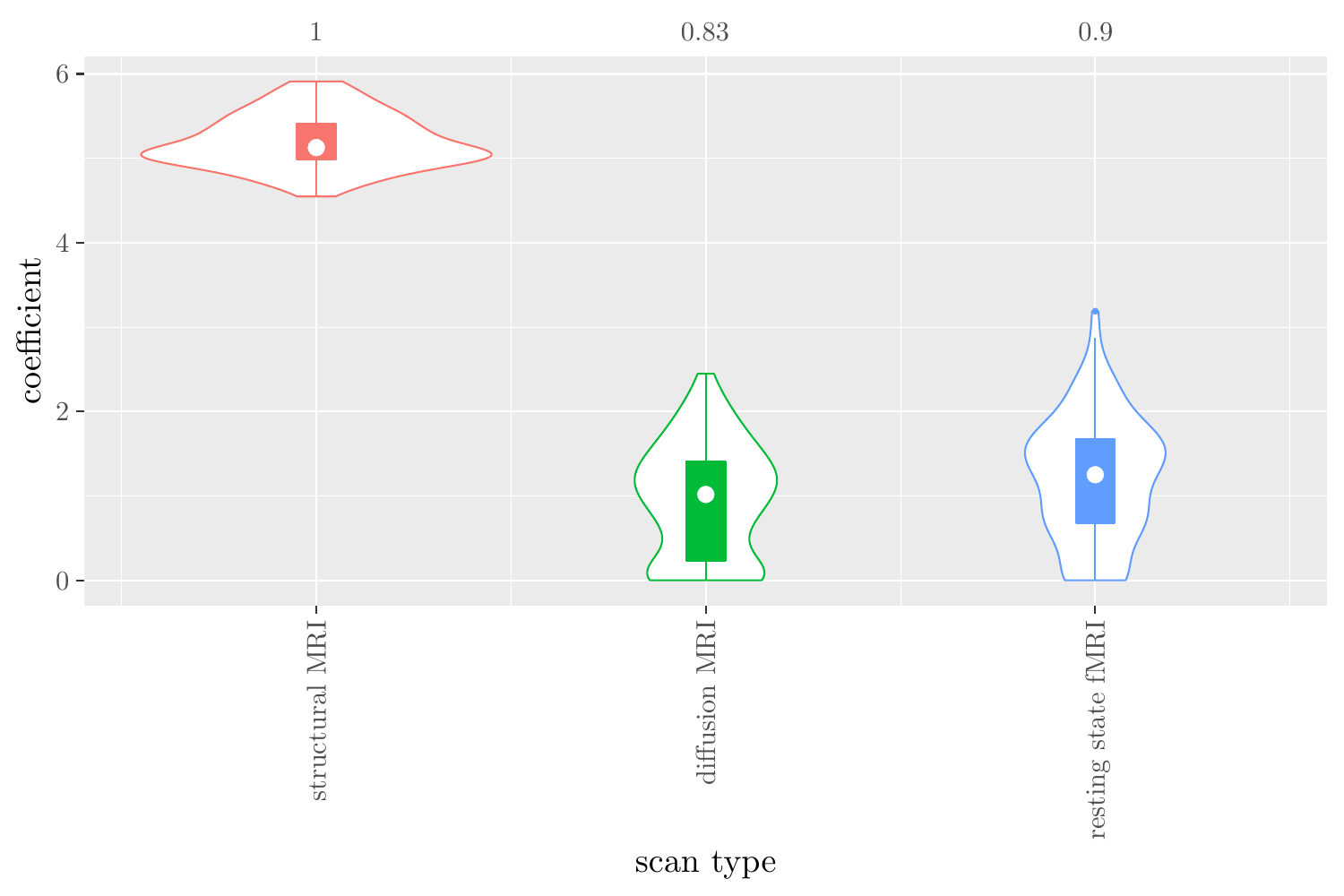}
	\caption{Box-and-violin plots of the hierarchical StaPLR meta-level regression coefficients for each scan type. The numbers at the top of the graph denote the proportion of times the coefficient was nonzero. \label{fig:meta_coef_plot}}
\end{figure}

\begin{figure}[]
	\centering
	\includegraphics{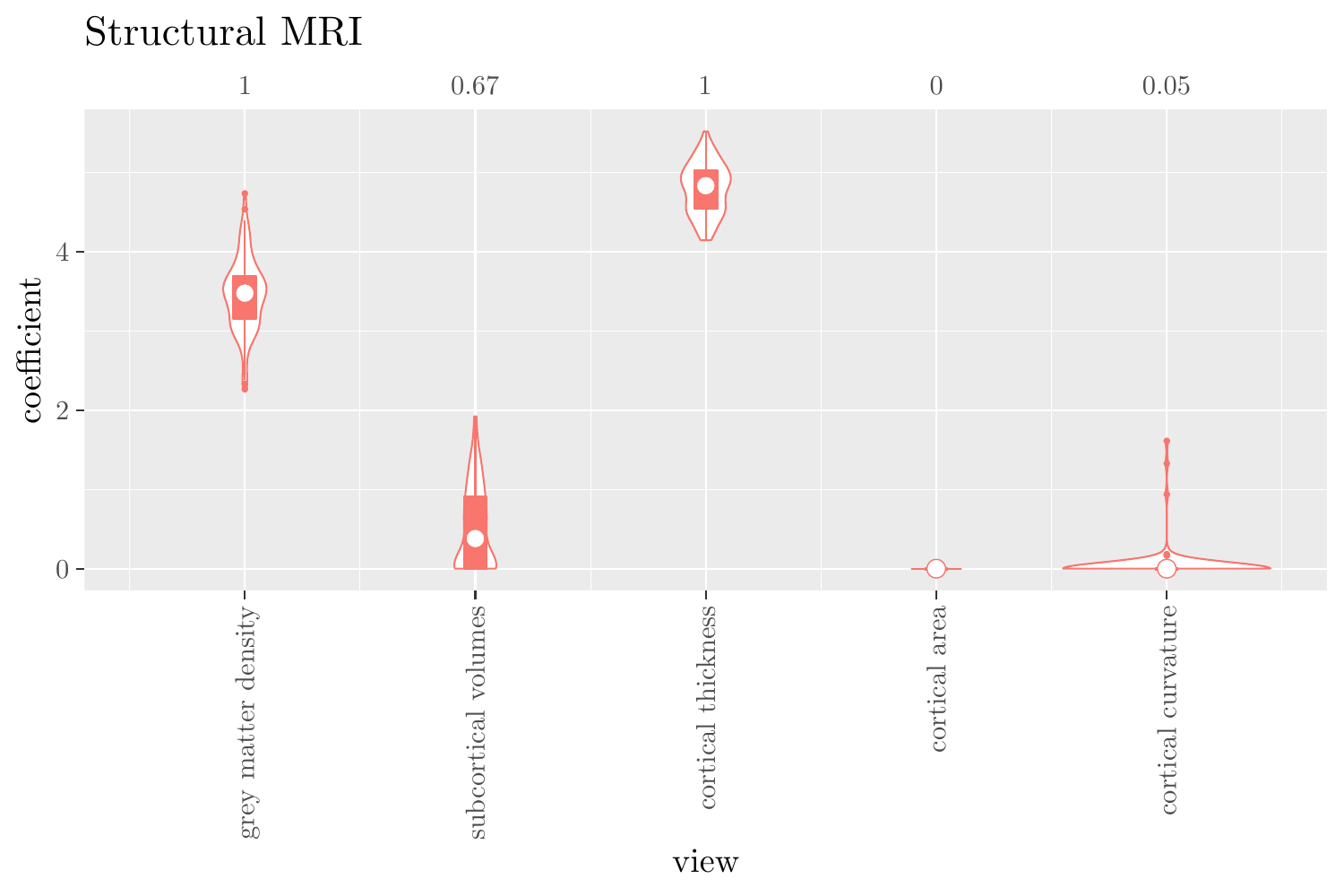}
	\caption{Box-and-violin plots of the hierarchical StaPLR intermediate-level regression coefficients for the structural scan type. The numbers at the top of the graph denote the proportion of times the coefficient was nonzero. \label{fig:struc_coef_plot}}
\end{figure}

\begin{figure}[]
	\centering
	\includegraphics{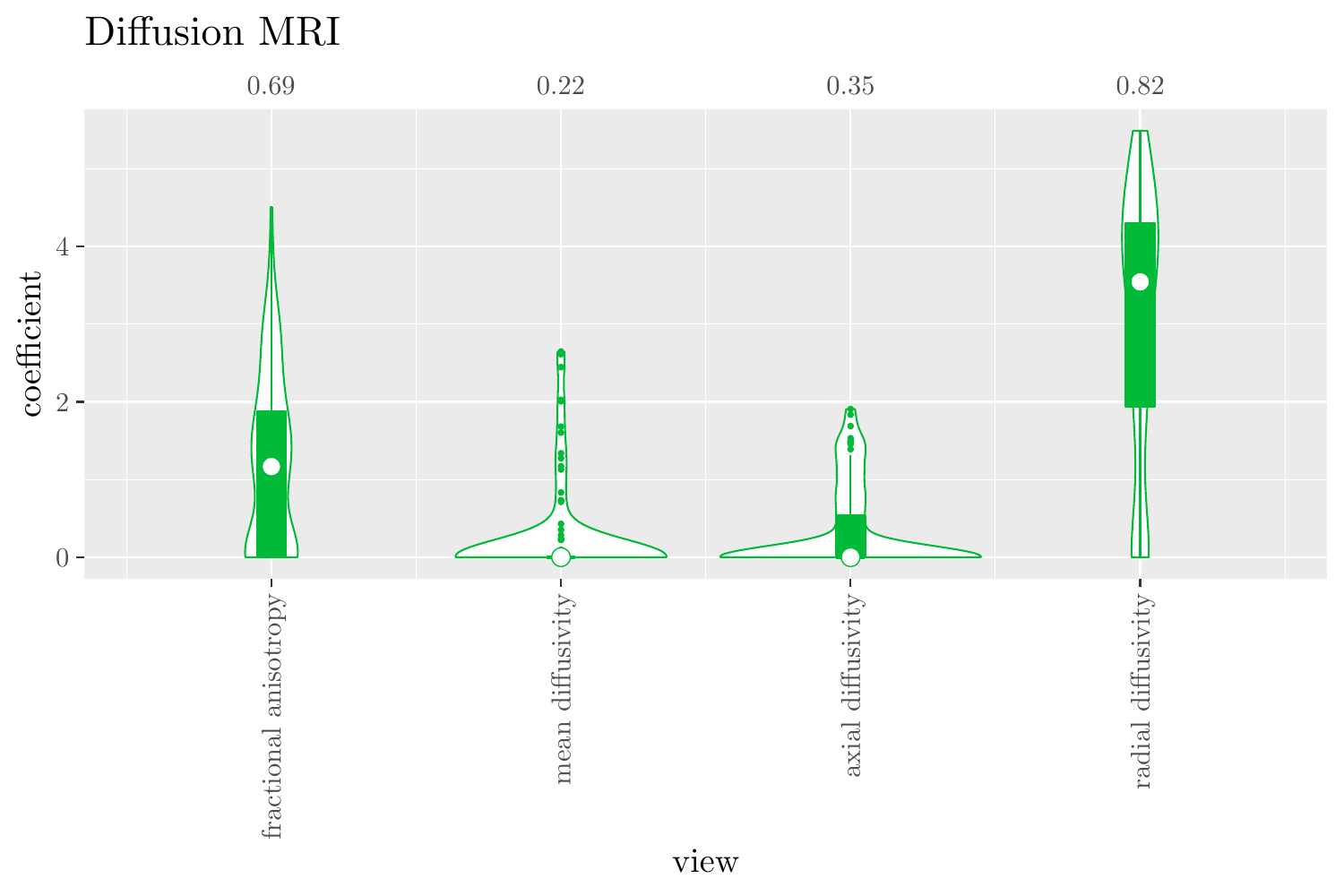}
	\caption{Box-and-violin plots of the hierarchical StaPLR intermediate-level regression coefficients for the diffusion scan type. The numbers at the top of the graph denote the proportion of times the coefficient was nonzero. \label{fig:dif_coef_plot}}
\end{figure}

\begin{figure}[]
	\centering
	\includegraphics{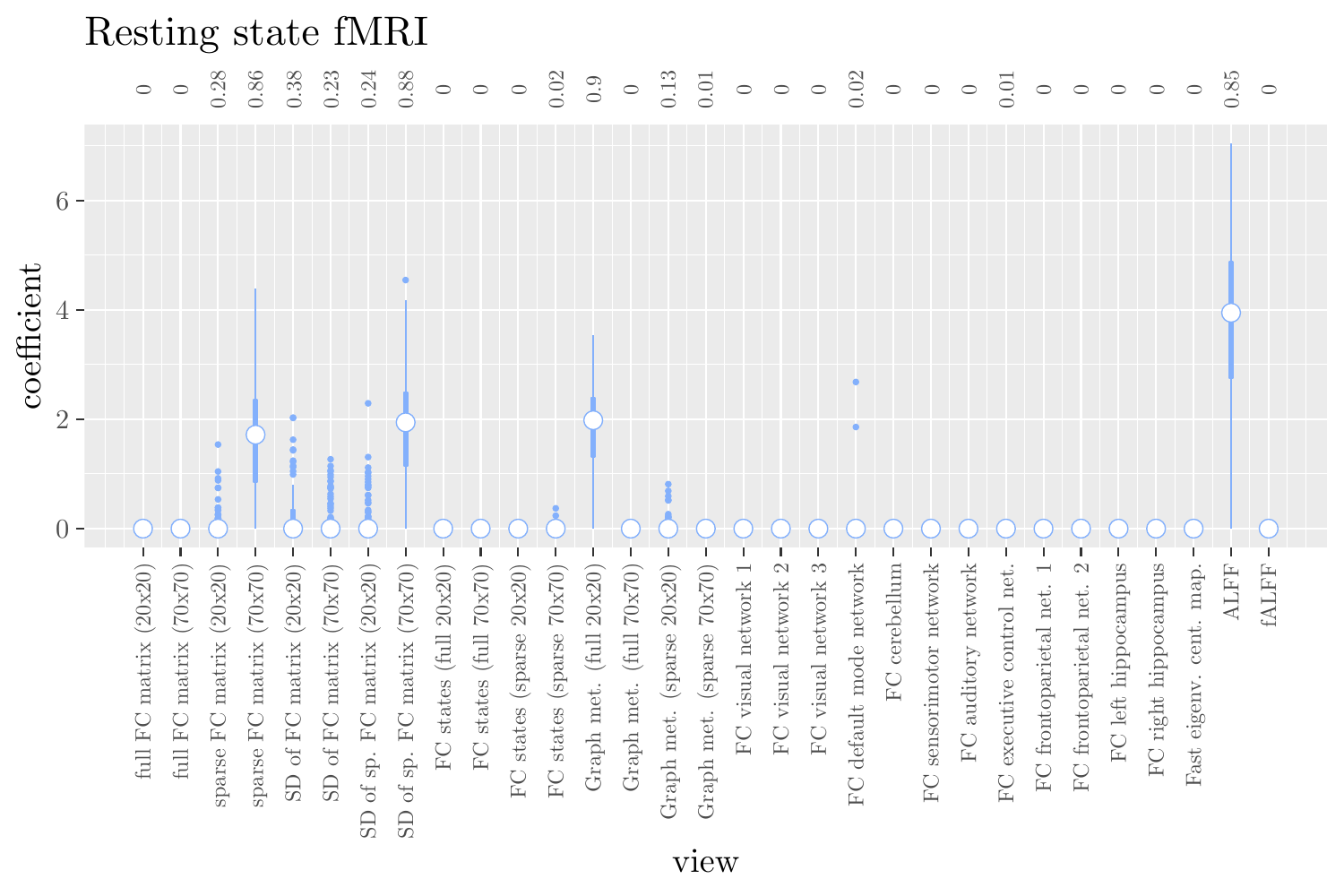}
	\caption{Box plots of the hierarchical StaPLR intermediate-level regression coefficients for the functional scan type. The numbers at the top of the graph denote the proportion of times the coefficient was nonzero. \label{fig:func_coef_plot}}
\end{figure}

The mean AUC of the hierarchical 3-level StaPLR was 0.942 (SD = 0.006), which was higher than that of the elastic net (0.922, SD = 0.008), and identical to that of the original StaPLR algorithm applied to either the MRI measures or the scan types. The mean accuracy was 0.893 (SD = 0.008), which was also higher than that of the elastic net (0.848, SD = 0.012), and comparable to that of the original StaPLR algorithm. Across the $10 \times 10$ fitted stacked classifiers, the structural scan was selected 100\% of the time, the diffusion weighted scan 83\% of the time, and the RS-fMRI scan 90\% of the time. The median regression coefficient for each scan type, as well as their distribution across the 10 $\times$ 10 fitted stacked classifiers can be observed in Figure \ref{fig:meta_coef_plot}. These are simply the regression coefficients in a logistic regression classifier. The input to this classifier is the output of the classifiers corresponding to each scan type, which are all predicted probabilities between zero and one. Taking the median values shown in Figure \ref{fig:meta_coef_plot}, the final predicted probability of Alzheimer's disease is given by an intercept plus 5.12 times the prediction from structural MRI, plus 1.02 times the prediction from diffusion-weighted MRI, plus 1.25 times the prediction from resting state fMRI. The final classification is thus largely determined by the classifier corresponding to the structural scan, with smaller contributions from the diffusion-weighted and resting state fMRI scans.
The contribution of each MRI measure within a given scan type can be compared in the same way. Figure \ref{fig:struc_coef_plot} shows that within the structural MRI scan type, the measures of cortical thickness and grey matter density contributed the most to the prediction. Subcortical volumes provided a much smaller contribution, and was not always selected. Cortical curvature was generally not selected and only provided a small contribution in 5\% of the fitted classifiers, while cortical area was never selected. Figures \ref{fig:dif_coef_plot} and  \ref{fig:func_coef_plot} show the contributions of the measures within the diffusion-weighted and resting state fMRI scan types, respectively. \par
\begin{figure}[]
	\centering
	\includegraphics{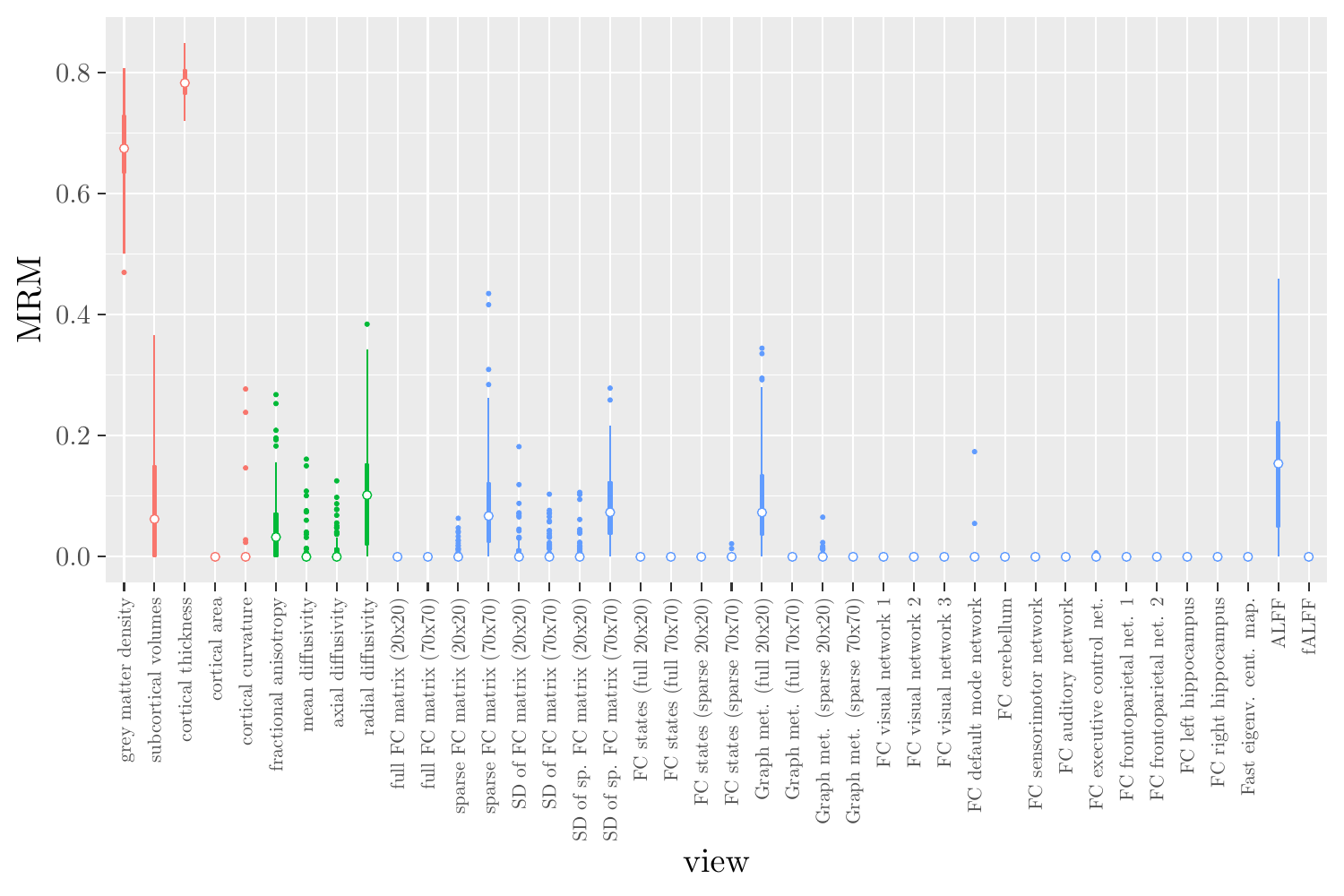}
	\caption{Influence of each of the MRI measures on the predictions of the hierarchical stacked classifier, as quantified by the minority report measure. MRM values range from 0 to 1, with 0 indicating no influence and 1 indicating maximum possible influence. The colors of the bars refer to the different scan types (red = structural MRI, green = diffusion MRI, blue = resting state fMRI). \label{fig:mrm_plot}}
\end{figure}

 
One important thing to consider when interpreting a StaPLR model with more than two levels, is that the coefficients shown in Figures \ref{fig:struc_coef_plot}, \ref{fig:dif_coef_plot}, and \ref{fig:func_coef_plot}, are coefficients of three different intermediate classifiers. Thus, we cannot simply compare coefficients across these figures. Doing so would lead us to conclude that ALFF (median coefficient of 4.05) is more important than grey matter density (median coefficient of 3.48). However, this would be an erroneous conclusion, since the structural scan type has a much larger weight than the resting state functional scan type (see Figure \ref{fig:meta_coef_plot}). To compare MRI measures across the different scan types we can use the minority report measure (MRM) introduced in Section \ref{sect:mrm}. Because the MRM measures the effect of the MRI measure-specific models on the final predicted outcome, it is suitable for comparing the importance of MRI measures even if they correspond to different scan types. We calculated the MRM for each measure, for each of the repetitions. As shown in Figure \ref{fig:mrm_plot}, the MRM properly reflects the high importance of the structural scan type compared with the diffusion and functional scan types. \par
If we compare the results of hierarchical StaPLR with the results of elastic net regression (Figures \ref{fig:elastic_plot} and \ref{fig:elastic_plot_norm}), we can observe both similarities and differences. In terms of the overall importance of the different scan types, the results are similar, with the structural MRI providing the MRI measures with the largest contribution, both in StaPLR (in terms of MRM and meta-level regression coefficient) and in the elastic net (in terms of the $L_2$ norm of the regression coefficients). In terms of the MRI measures within the structural scan type, the results are also similar, with cortical thickness being the most important measure, followed by grey matter density, and subcortical volumes. The fact that both methods agree on the same MRI measures being the most important for the classification of Alzheimer's disease provides somewhat of a `sanity check'. Within the scan types which have a smaller contribution, i.e. diffusion-weighted MRI and resting state fMRI, we can see differences between the methods. For example, in StaPLR mean diffusivity is not considered important, while it is of some importance in the elastic net model. The largest differences, however, are seen within the functional scan type. StaPLR generally included only 4 resting state fMRI measures, whereas elastic net generally included features from 17 fMRI measures. Features from ALFF are included by both methods. Although the StaPLR model is much sparser in terms of the MRI measures which are included, this did not lead to a reduction in accuracy. In fact, the accuracy of the StaPLR model compares favorably to that of the elastic net.

\FloatBarrier

\section{Discussion} 

We have extended the StaPLR algorithm to adapt to a hierarchical multi-view data structure. That is, the extended StaPLR algorithm allows for the analysis of datasets containing large numbers of features which are nested within lower-level views, which are in turn nested within higher-level views. We applied this extension to a multi-view MRI data set in the context of Alzheimer’s disease classification, where features are nested within MRI measures, which are in turn nested within scan types. The presented application can serve as an example of a more general class of applications within neuroimaging and the biomedical sciences. In our specific application to AD classification, the classifier produced by StaPLR was more accurate than the one produced by elastic net regression. 
We have shown how in StaPLR the relative importance of MRI measures derived from the same scan type can easily be compared using their regression coefficient. Additionally, we have introduced the minority report measure, which allows for comparing the importance of MRI measures derived from different scan types. \par 
If we compare the results of the hierarchical extension of StaPLR with the original algorithm applied to either MRI measures or scan types, we see three different resulting classifiers which have nearly identical classification performance. Naturally, the results of the StaPLR algorithm depend on the specified multi-view structure, and specifying a different structure will lead to a different model. Since MRI data generally contain high amounts of collinearity it is not all that surprising that different models may have similar classification performance. However, it is still natural to ask which of these three models is the `best'. Since all three models have nearly identical classification performance, and statistical models are generally `wrong' in the sense that they can only provide a simplification or approximation of reality, the most important question is probably: which approach is the most useful? We argue that the hierarchical StaPLR model is more useful for several reasons. First, hierarchical StaPLR is the only approach that accurately matches the study design because it most closely follows how the data are collected and processed: data are collected through different MRI scans, and then researches derive different MRI measures from these scan types. Second, unlike the original StaPLR algorithm, hierarchical StaPLR allows easy computation of measures of importance for both MRI measures and scan types. If applying the original StaPLR algorithm, two separate analyses are required, which considerably increases computational cost. In addition, the two different analyses may lead to different conclusions, as was the case in our data set (compare Figure \ref{fig:MVS_bottom_coef_plot} with \ref{fig:MVS_top_coef_plot}). If the MRI measures are used as views, some resting state fMRI measures are generally included, but if the scan types are used instead, the fMRI scan type is always discarded. This discrepancy is probably caused by the fact that when the scan types are used as views, the algorithm is forced to select or discard all features within this scan type. Since the fMRI scan type has a very high number of features and likely a low signal-to-noise ratio, the addition of a large amount of noise to the model is not worth the inclusion of the scan type by the two-level StaPLR algorithm. However, if a selection of MRI measures is made first, such as in the hierarchical StaPLR model, then signal present in the fMRI scan type can still be picked up by the algorithm. \par
Given the results of the hierarchical StaPLR algorithm, we can compare the relative importance of the different MRI measures using the regression coefficients or the MRM. However, we may additionally want to make a binary decision: is this measure required for prediction of the outcome or not? This is of course more difficult, since although some measures were selected 100\% or 0\% of the time, for many measures the situation is not so clear-cut. One approach would be to say that for an MRI measure to be important, it should have been selected at least 50\% of the time (i.e. its median coefficient should be nonzero). In this case we would select three structural measures (cortical thickness, grey matter density and subcortical volumes), two diffusion measures (fractional anisotropy and radial diffusivity), and four functional measures (ALFF, the graph metrics as computed from the full 20x20 FC matrix, the sparse 70x70 FC matrix, and the SDs associated with the latter), for a total of nine selected MRI measures. Note that this is considerably sparser than the elastic net, for which the selected features were on average spread out over 24 MRI measures. 
It is also interesting to see that the observed selection probabilities for the different MRI measures are not generally in the neighborhood of 50\%. Instead, all measures were included either at least 67\% of the time, or less than 38\% of the time, providing a clear separation into a ``frequently selected" and ``infrequently selected" group of MRI measures. \par 
Of course, binary decisions regarding which MRI measures or scan types are required for prediction are further complicated by the fact that we have shown different two-level StaPLR models with comparable classification performance. However, it is important to note that these models correspond to different research questions. For example, when using only the scan types as views, the research question pertains to the relevance of \textit{complete} scan types (i.e. including any noise), rather than to the most informative subset of MRI measures derived from those scan types, as in the hierarchical StaPLR model. Naturally, a different research question may lead to a different answer. \par 
The results of hierarchical StaPLR model are in line with earlier research. Hierarchical StaPLR, the original StaPLR algorithm using MRI measures as views, and the elastic net all appeared to agree on the structural MRI measures of grey matter density and cortical thickness being the most important for classification, which is in line with earlier research identifying measures of grey matter atrophy as important bio-markers for Alzheimer's disease \citep{Lerch2005, Frisoni2010}. The largest difference between the methods was seen in terms of fMRI measures, of which hierarchical StaPLR selected 4, StaPLR using only the MRI measures selected 2, StaPLR using only the scan types selected none, and the elastic net selected 17. In particular, the elastic net appeared to include more features from the larger feature sets (Figure \ref{fig:elastic_plot}), such as the feature sets containing the voxel-wise functional connectivity with individual RSNs. In contrast, hierarchical StaPLR includes MRI measures which contain summarizing information about RSNs (i.e. graph metrics, the sparse 70x70 FC matrix, and the dynamics of the sparse 70x70 FC matrix). The importance of the 70x70 FC matrix and its dynamics are in line with the results of a previous study which used only the resting state fMRI scans for AD classification \citep{deVos2017}.
The results of the hierarchical StaPLR analysis suggest that although the structural scan type is dominant in the classification of Alzheimer's disease, diffusion MRI and resting-state fMRI can both provide useful contributions to the classification. These results are broadly in line with a previous study investigating the relevance of a smaller subset of structural, diffusion and resting-state functional MRI measures \citep{Schouten2016}. \par 
Although hierarchical StaPLR selected all MRI scan types, it did make a smaller selection of required MRI measures. In practice, such a selection may translate to less time spend on the computation of different feature sets from MRI scans. Furthermore, our results indicate that StaPLR can adequately deal with imbalanced numbers of features within each view. Whereas standard elastic net tends to select many features from very large views (e.g., the functional connectivity views; Figure \ref{fig:elastic_plot}), StaPLR does not show this preference for large views (most functional connectivity views obtained a weight of 0 as shown in Figure \ref{fig:func_coef_plot}). A drawback of StaPLR, which it shares with all penalized regression methods, is that for any single run of the algorithm the selection is binary: a view is either selected or not. As discussed above, the actual set of selected views may vary from run to run. In this article, we have quantified this variability by showing the distribution of results over all repeated cross-validation folds. Other re-sampling methods, such as the bootstrap, could also be used to gain more insight in the stability of the results. However, compared with subsampling, bootstrapping may increase the likelihood of noise variables being selected \citep{deBin2016}. In addition, re-sampling methods are typically computationally expensive. In the future, we therefore aim to introduce a form of uncertainty quantification, such as confidence intervals, that can be computed from only a single run of the StaPLR algorithm. \par
As mentioned before, the results of the StaPLR algorithm depend on the specified multi-view structure. In our analysis, features were nested in \textit{MRI measures}, which were in turn nested in \textit{scan types}. The multi-view structure was specified this way because it matches the study design. However, one could specify a different multi-view structure to match a different research question. In fact, we have done so when applying StaPLR with only two levels. Another example of a different research question would occur if the primary interest is in identifying which \textit{brain areas} are the most important for AD classification. In this case, one could treat each brain area as a separate view. This may, of course, again lead to different results. For example, the feature set that consists of the volumes of the subcortical structures was found to play only a minor role in AD classification in the hierarchical StaPLR model, whereas this feature set also contains the volumes of the left and right hippocampus that are considered to be AD hallmarks. Decoupling the volumes of the different subcortical structures and treating each brain area as a separate view would allow each structure to obtain its own weight. In such an analysis, one might see an increased importance of certain structures traditionally associated with AD, such as the hippocampus. However, such an analysis is outside the scope of this article. \par
The application shown in this article serves as an example of how StaPLR can be applied to hierarchical multi-view data. It should be noted that the method can be further extended to a more complex structure, such as a hierarchical structure with more levels, or a structure with a mixed number of levels. The latter may be of particular importance when the data is collected from entirely different domains. For example, the hierarchical multi-view structure for MRI data may be quite different from that of genetic data, other biomarkers, or clinical variables. Such a difference can easily be handled by the StaPLR algorithm, paving the way for applications to larger multi-source data sets such as those obtained through the UK Biobank initiative. 

\section{Conclusion}

We have extended the StaPLR algorithm to hierarchical multi-view MRI data, and applied it to Alzheimer's disease classification. We have shown that StaPLR produces a stacked classifier that allows researchers to see which scan types, and which MRI measures derived from those scan types, play the most important role in classification. In addition, the stacked classifier showed an increase in classification accuracy when compared with logistic elastic net regression.

\newpage

\section*{Conflict of interest statement}

The authors declare that the research was conducted in the absence of any commercial or financial relationships that could be construed as a potential conflict of interest.

\section*{Author contributions}

\textbf{Wouter van Loon:} Conceptualization, Methodology, Software, Formal analysis, Investigation, Writing - original draft, Writing - review \& editing, Visualization. \\*
\textbf{Frank de Vos:} Investigation, Data curation, Writing - review \& editing. \\*
\textbf{Marjolein Fokkema:} Writing - review \& editing. \\*
\textbf{Botond Szabo:} Writing - review \& editing. \\* 
\textbf{Marisa Koini:} Data curation. \\*
\textbf{Reinhold Schmidt:} Data curation. \\* 
\textbf{Mark de Rooij:} Conceptualization, Writing - review \& editing, Supervision.

\section*{Funding}

This research was funded by Leiden University. Botond Szabo received funding from the European Research Council (ERC) under the European Union’s Horizon 2020 research and innovation programme (grant agreement No 101041064).

\section*{Ethics statement}

The original PRODEM study was approved by the ethics committees of the Medical University of Graz, the Medical University of Innsbruck, the Medical University of Vienna, the Konventhospital Barmherzige Brüder Linz, the Province of Upper Austria, the Province of Lower Austria and the Province of Carinthia \citep{PRODEM}. Written informed consent was obtained from all patients and their caregivers \citep{PRODEM}. For collection of the original ASPS data, standard protocol approvals, registrations, and patient consents were obtained \citep{ASPS2}. The study was approved by the standard ethics committee of the Medical University of Graz for experiments using human participants, and written informed consent was obtained from all study participants \citep{ASPS2}.

\section*{Data availability statement}
The source code of the R package `multiview' is publicly available \citep{multiview}, as are the R scripts used for model fitting and evaluation \citep{code_repo}. The MRI data used in this study is available upon direct request to the 6th author (R. Schmidt); a formal data sharing agreement is mandatory.

\newpage

\appendix

\section{Supplementary materials}

These supplementary materials describe the process of obtaining the features used in ``Analyzing hierarchical multi-view MRI data with StaPLR: An application to Alzheimer's disease classification". This appendix is largely a reiteration of previous work, in particular that of \citet{deVos2016,deVos2017} and \citet{Schouten2016, Schouten2017}. The relevant information from these publications is collected here for the reader's convenience.

\subsection{Participants}

The data were collected at the Medical University of Graz in Austria, and consisted of 76 clinically diagnosed probable AD patients and 173 cognitively normal elderly. The AD patients were part of the prospective registry on dementia (PRODEM) \citep{PRODEM}. The inclusion criteria for PRODEM are: Dementia diagnosis according to DSM-IV criteria \citep{DSM}, AD diagnosis according to the NINCDS-ADRDA criteria \citep{McKhann2011}, non-institutionalisation or need for 24-h care, and the availability of a caregiver who agrees to provide information on the patients’ and his or her own condition. Patients were excluded if co-morbidities were likely to preclude successful completion of the study. Informed consent was obtained from all patients and their caregivers. We only included patients for which anatomical MRI, diffusion MRI and rs-fMRI were available. The controls were scanned at the same scanning site, over the same period, with the same scanning protocol as the AD patients as a part of the Austrian Stroke Prevention Study (ASPS). The ASPS is a community-based cohort study on the effects of vascular risk factors on brain structure and function in elderly participants without a history or signs of stroke and dementia on the inhabitants of Graz, Austria \citep{ASPS1, ASPS2}. Informed consent was obtained from all participants.

\subsection{MRI acquisition}

Each participant was scanned on a Siemens Magnetom TrioTim 3 T MRI scanner. Anatomical T1-weighted images were acquired with TR = 1900 ms, TE = 2.19 ms, flip angle = 9, 179 slices, with an isotropic voxel size of 1 mm. \par 
Diffusion images were acquired along 12 non-collinear directions, scanning each direction 4 times with TR = 6700 ms, TE = 95 ms, 50 axial slices, voxel size = $2.0 \times 2.0 \times 2.5$ mm. \par 
Resting-state fMRI series of 150 volumes were obtained with TR = 3000 ms, TE = 30 ms, flip angle = 90, 40 axial slices, with an isotropic voxel size of 3 mm. Participants were instructed to lie still with their eyes closed, and to stay awake.

\subsection{MRI preprocessing} 

The MRI data of all subjects were preprocessed using the FMRIB Software Library (FSL version 5.0) \citep{Jenkinson2012, Smith2004}. For the anatomical MRI scans, we applied brain extraction and bias field correction. For the diffusion MRI scans, we applied brain extraction and eddy current correction. For the rs-fMRI data, this included brain extraction, motion correction, a temporal high pass filter with a cutoff point of 100 seconds, 3 mm FWHM spatial smoothing, and non-linear registration to standard MNI152 space. Additionally, we used ICA-AROMA to automatically identify and remove noise components from the fMRI time course \citep{Pruim2015}. ICA-AROMA adequately removes motion related noise from fMRI data, without the need for removing volumes with excessive motion \citep{Parkes2018}. 

\subsection{Feature extraction}

\subsubsection{Structural MRI}

The process of extracting the features corresponding to cortical thickness, area, curvature, grey matter density, and subcortical volumes, is identical to the process described in \citet{deVos2016}. For completeness, we also describe it below. \par
In order to calculate cortical thickness, cortical area and cortical curvature, the raw (not preprocessed) T1-weighted images were processed using Freesurfer 5.3.0 \citep{Dale1999, Fischl1999}. First, this entails intensity normalization and brain extraction \citep{deVos2016}. Using the resulting image the boundary between grey and white matter was located, and a triangular mesh was constructed around the white matter surface \citep{deVos2016}. The grey matter surface was created by deforming the mesh outward so that it closely followed the boundary between grey matter and cerebral spinal fluid \citep{deVos2016}. Cortical thickness was calculated as the distance between the white matter and grey matter surface for each vertex \citep{deVos2016}. The image was then registered to the Freesurfer common template using the image's cortical folding pattern \citep{deVos2016}. The neocortex was parcellated into the 68 regions of the Desikan-Killiany atlas \citep{Desikan2006}. The thickness of each parcellation unit was calculated as the mean thickness of all the vertices within that parcellation \citep{deVos2016}. Thus, 68 cortical thickness features are obtained per subject. The cortical surface area was calculated by summing the areas of the grey matter mesh triangles for each parcellation, yielding 68 cortical area features per subject \citep{deVos2016}. To obtain the cortical curvature features, the mean of the curvature values in the two principal directions of the of the surface was calculated \citep{deVos2016}. The curvature of a vertex in these directions was calculated as the inverse of the length of the radius of osculating circles in these directions \citep{Ronan2011}. The curvature values of the vertices were averaged for each of the parcellations, yielding 68 cortical curvature features per subject \citep{deVos2016}. \par
Grey matter density was calculated using FSL VBM (version 5.0.7) \citep{Ashburner2000, Smith2004}. The brain-extracted images were first segmented into grey matter, white matter, and CSF \citep{deVos2016}. A study-specific grey matter template was created in two steps. First, the grey matter images were affine-registered to the ICBM-152 grey matter template and the resulting images were averaged to create a first-pass template \citep{deVos2016}. Then, the grey matter images were nonlinearly registered to the first-pass temlate and the resulting images were averaged to obtain a final template at $2 \times 2 \times 2 \text{mm}^3$ resolution in standard space \citep{deVos2016}. The grey matter images were then registered to the final template and smoothed with a Gaussian kernel with a full width at half maximum of 3 mm \citep{deVos2016}. The voxel wise values were then averaged within the 48 regions of the probabilistic Harvard-Oxford cortical atlas \citep{deVos2016}. The 48 grey matter density features were obtained by calculating the weighted averages of the regions, with voxels contribution to the average of a region based on their probability of being part of that region \citep{deVos2016}. \par
The volumes of the subcortical structures were calculated using the FMRIB's Integrated Registration and Segmentation Tool (FIRST) in FSL \citep{Patenaude2011}. The whole-head images were affine registered to the nonlinear MNI-152 template \citep{deVos2016}. In a second stage, initialized by the result of the first stage, a subcortical mask was used to achieve a more accurate and robust affine registration \citep{deVos2016}. The shapes of the subcortical structures were modeled by deformable meshes and the boundary voxels were classified as being part of the subcortical structure using structural segmentation \citep{Zhang2001}. The cortical volumes were then corrected for intracranial volume as obtained by FSL, yielding 14 subcortical volume features per subject, corresponding to the thalamus, caudate, putamen, pallidum, hippocampus, amygdala and accumbens of both hemispheres \citep{deVos2016}.  

\subsubsection{Diffusion-weighted MRI}

The diffusion MRI scans were used to calculate fractional anisotropy (FA), mean diffusivity (MD), axial diffusivity (DA), and radial diffusivity (DR). First, DTIFIT in FSL \citep{Jenkinson2012, Smith2004} was used to fit a diffusion tensor model at each voxel to calculate voxel-wise FA, MD, DA and DR images for each subject. Then subjects’ FA, MD, DA and DR images were projected onto the FMRIB58\_FA mean FA image using tract based spatial statistics (TBSS) \citep{Smith2006}. Finally,  weighted averages of the FA, MD, DA and DR values were calculated within the 20 regions of the probabilistic JHU white-matter tractography atlas \citep{Hua2008}, yielding 20 features for FA as well as MD, DA and DR \citep{Schouten2016}.

\subsubsection{Resting state fMRI}

The resting state fMRI feature sets used in this article have already been described in detail in \citet{deVos2017}; the following is a reiteration of the most relevant sections. \par 
Resting state networks (RSNs) were obtained using temporal concatenation independent component analysis (ICA) in FSL MELODIC \citep{MELODIC}. The functional data of all participants was registered to standard space and concatenated along the time dimension. ICA was performed on the concatenated dataset, once with 20 and once with 70 components \citep{deVos2017}. The resulting ICA component weight maps were registered back to subject space, weighted by subject specific grey matter density maps, and multiplied with the functional data, resulting in mean time course for each component \citep{deVos2017}. These time course were then used to calculate functional connectivity matrices using both full and sparse partial correlations \citep{deVos2017}. The partial correlation matrices were calculated using the graphical lasso \citep{graphical_lasso} implemented in MATLAB \citep{MATLAB}, with $\lambda = 100$ \citep{deVos2017}. The resulting two 20 $\times$ 20 matrices contain 190 unique elements, and the 70 $\times$ 70 matrices contain 2415 unique elements to be used as candidate features in the classification. Any features with zero variance were removed. \par
The dynamics of the FC matrices were calculated using a sliding window approach with a window size of 33s \citep{deVos2017}. The windows were shifted one volume at a time, leading to 140 windows \citep{deVos2017}. The previously described four FC matrices were calculated within each window, and the standard deviation of the FC matrices of all windows was obtained \citep{deVos2017}. \par
The sliding window FC matrices were clustered using k-means clustering (with k=5 and Manhattan distance) to obtain 5 "FC states" \citep{deVos2017}. The number of sliding window matrices that were assigned to each of the five FC states was then calculated for each participant \citep{deVos2017}. \par
Graph metrics were calculated using the Brain Connectivity Toolbox \citep{Rubinov2010} in MATLAB \citep{MATLAB} for each of the four FC matrices. Connection strength, weighted betweenness centrality and weighted clustering coefficients were calculated for every node, and weighted characteristic path length and weighted transitivity for the entire network \citep{deVos2017}. Additionally, several graph metrics were calculated on binarized versions of the FC matrices: connection degree, betweenness centrality and clustering coefficient for every node, and characteristic path length and transitivity for the entire network \citep{deVos2017} \par
Whole brain FC with 10 RSNs was calculated using dual regression in FSL \citep{Filippini2009}, using the RSN templates of \citet{Smith2012}. The voxel-wise whole brain FC results for each of the 10 RSNs were used as 10 distinct feature sets. Voxel-wise whole brain FC maps were additionally obtained for the left and right hippocampus \citep{deVos2017}. \par 
Eigenvector centrality maps were calculated for each participant using fastECM \citep{Wink2012,Binnewijzend2014}.
The amplitude of low frequency fluctuations (ALFF) \citep{Yufeng2007, Biswal2010} and fractional ALFF \citep{Zou2008} were calculated for each participant using REST \citep{REST}. The voxels' ALFF and fALFF values were divided by the mean ALFF/fALFF within a subjects whole brain \citep{Zou2008, deVos2017}.

\bibliographystyle{IEEEtranN}
\bibliography{Bibliography} 

\end{document}